\documentclass[prd,amssymb,10pt]{revtex4}
\usepackage{amsmath,amsfonts,amssymb}
\usepackage{verbatim,graphicx,float}

\begin{document}

\title{Relic gravitons from super-inflation}

\author{Jakub Mielczarek}
\email{jakubm@poczta.onet.pl}
\affiliation{\it Astronomical Observatory, Jagiellonian University, 30-244
Krak\'ow, Orla 171, Poland}
\affiliation{\it The Niels Bohr Institute, Copenhagen University, Blegdamsvej 17, 
DK-2100 Copenhagen, Denmark}

\author{Marek Szyd{\l}owski}
\email{uoszydlo@cyf-kr.edu.pl}
\affiliation{Department of Theoretical Physics, 
Catholic University of Lublin, Al. Rac{\l}awickie 14, 20-950 Lublin, Poland}
\affiliation{Marc Kac Complex Systems Research Centre, Jagiellonian University,
Reymonta 4, 30-059 Krak{\'o}w, Poland}


\begin{abstract}
The super-inflationary phase is predicted by the Loop Quantum Cosmology.
In this paper we study the creation of gravitational waves during this phase.
We consider the inverse volume corrections to the equation for the tensor modes
and calculate the spectrum of the produced gravitons. The amplitude of the 
obtained spectrum as well as maximal energy of gravitons strongly depend on the 
evolution of the Universe after the super-inflation. We show that a further 
standard inflationary phase is necessary to lower the amount of gravitons below 
the present bound. In case of the lack of the standard inflationary phase, the 
present intensity of gravitons would be extremely large. These considerations 
give us another motivation to introduce the standard phase of inflation. 
\end{abstract}

\maketitle

\section{Introduction} \label{sec:intro}

The cosmological creation of the gravitational waves was proposed by Grishchuk \cite{Grishchuk:1974ny} 
in the mid-seventies. Since that time this phenomenon has been studied extensively, especially in
the context of the inflation. The accelerating expansion phase gives the conditions for the abundant 
creation of the gravitational waves. Gravitons produced during the inflation fill the entire space in the
form of a stochastic background. Together with the scalar modes, produced during the inflation, they form 
primordial perturbations leading to the structure formation. The analysis of the cosmic microwave background (CMB) and large scale
structures gives therefore the possibility of testing inflationary models. In the case of the CMB the impact 
of the gravitational waves comes from the primordial spectrum and from tensor Sachs-Wolfe effects. 
The Sachs-Wolfe effect is somehow secondary and leads to the CBM anisotropies as the result of the scattering of 
CMB photons on the relic gravitons. The form of this anisotropies is given by
\begin{equation}
\left( \frac{\Delta \text{T}}{\text{T}}  \right)_{\text{t}} =
-\frac{1}{2} \int_{\tau_1}^{\tau_{2}} d \tau \ h'_{ij} n^i n^j  
\end{equation}
where $h_{ij}$ describes tensor modes and $n^i$ is the vector parallel to the unperturbed geodesics. 
The influence of the gravitational waves for the CMB is however to weak to be observed 
directly with the present observational abilities. Another possible method to 
detect gravitational waves is to use of the antennas like LIGO, VIRGO, TAMA or GEO600
\cite{Abbott:2003vs,Cella:2007jh} .
Although these detectors are now very sensitive this is still not enough 
to detect directly the gravitational waves background \cite{Abbott:2007wd}. It may look pessimistic,
we hope however that some further 
improvement of the observational skills bring us the observational evidence, so needful
for the further theoretical improvements. 

In this paper we consider a new type of the inflation which naturally occurs in the Loop
Quantum Cosmology \cite{Bojowald:2006da}.
This is so called the super-inflationary scenario \cite{Bojowald:2002nz,Copeland:2007qt}
and is a result of the quantum nature of spacetime in the Planck scales. 
The spacetime is namely discrete in the quantum regime and its  
evolution is governed by discrete equations. However for the scales greater than $a_i=\sqrt{\gamma}l_{\text{Pl}}$
($\gamma $ is so called Barbero-Immirzi parameter)  
the evolution of the spacetime can be described by the Einstein equations with quantum corrections.
For typical values of the quantum numbers, super-inflationary phase takes place in this semi-classical region.

Our goal is to describe the production of the gravitational waves during the super-inflation. This problem 
was preliminary analysed in Ref.~\cite{Mielczarek:2007zy}, but quantum corrections to the equation for tensor modes 
was not included to calculate the spectrum of the gravitons. In this paper we include so called inverse
volume corrections to the equation for evolution of the tensor modes and then calculate the spectrum 
of produced gravitational waves. The equations for the tensor modes 
was recently derived by Bojowald and Hossain \cite{Bojowald:2007cd}. They had analysed the inverse volume corrections and
corrections from holonomies. In this paper we concentrate on these first ones. 
The quantum corrections are generally complicated functions but they have simple asymptotic behaviours.
To calculate the productions of gravitons during some process we need somehow to know only initial 
and final states, where asymptotic solutions are good approximation. In these regimes
calculations can be done analytically. We use numerical solutions to match them. 

The organization of the text is the following. In section II we fix the 
semi-classical dynamics. Then in section III we consider creation of the 
gravitons on the defined background. In section IV we summarize the results.

\section{Background dynamics}

The formulation of Loop Quantum Gravity bases on the Ashtekar variables \cite{Ashtekar:1987gu} and holonomies. The Ashtekar
variables replace the spatial metric field $q_{ab}$ in the canonical formulation as follow 
\begin{eqnarray}
A^i_a &=& \Gamma^i_a+\gamma K_a^i  , \\
E^a_i &=& \sqrt{|\det q|} e^{a}_i
\end{eqnarray}  
where $\Gamma^i_a$ is the spin connection defined as
\begin{equation}
\Gamma^i_a = -\epsilon^{ijk}e^b_j(\partial_{[a}e^k_{b]}+\frac{1}{2}e^c_k e^l_a \partial_{[c}e^l_{b]} )
\end{equation}
and the $K_a^i$ is the intrinsic curvature. The $e^{a}_i$ is the inverse of the co-triad $e^i_a$ defined as $q_{ab}=e_a^ie_b^j$. 
In terms of the Ashtekar variables the full Hamiltonian for general relativity is a sum of constraints
\begin{equation}
H_{\text{G}}^{\text{tot}}= \int d^3 {\bf x} \, (N^i G_i + N^a C_a + N  h_{\text{sc}}),
\end{equation}
where 
\begin{align}
C_a &= E^b_i F^i_{ab} - (1-\gamma^2)K^i_a G_i ,\nonumber \\
G_i &= D_a E^a_i
\end{align}
and the scalar constraint has a form
\begin{align}\label{ham}
&H_{\text{G}}:=\int d^3{\bf x} \, N(x) h_{\rm sc}= \nonumber \\ 
  &\frac{1}{16 \pi G} \int d^3{\bf x} \, N(x)\left( \frac{E^a_i
  E^b_j}{\sqrt{|\det E|}} {\varepsilon^{ij}}_k F_{ab}^k -
  2(1+\gamma^2) \frac{E^a_i E^b_j}{\sqrt{|\det E|}} K^i_a
  K^j_b \right) 
\end{align}
with $F=dA + \frac{1}{2}[A,A]$. The full Hamiltonian of theory is a sum of the gravitational and matter part.
With convenience as a matter part we choose the scalar field with the Hamiltonian
\begin{equation}
H_{\phi}=\int d^3{\bf x} \, N(x)\left( \frac{1}{2}\frac{\pi^2_{\phi}}{\sqrt{|\det E|}}  + \frac{1}{2} \frac{E^a_i
  E^b_i \partial_a \phi \partial_b \phi  }{\sqrt{|\det E|}}  + \sqrt{|\det E|} V(\phi) \right).
\end{equation}
We assume here that field $\phi$ is homogeneous and 
start his evolution from the minimum of potential $ V(\phi)$. The second assumption
states that contribution from potential term is initially negligible. So the density of Hamiltonian $H_{\phi}$ is simplified to 
the form $\mathcal{H}_{\phi}=(1/2)\pi^2_{\phi}/\sqrt{|\det E|}$. The term $1/\sqrt{|\det E|}$ for the classical
FRW universe corresponds to $1/a^3$ where $a$ is the scale factor. On the quantum level term $1/\sqrt{|\det E|}$
is  quantised and have discrete spectrum. In the regime 
$a \gg a_i$ we can however use the approximation $1/\sqrt{|\det E|}=D/a^3$  where  
\begin{equation}
D(q)=q^{3/2} \left\{ \frac{3}{2l} \left(  \frac{1}{l+2}\left[(q+1)^{l+2}-|q-1|^{l+2} \right]-
\frac{q}{1+l}\left[(q+1)^{l+1}-\mbox{sgn}(q-1)|q-1|^{l+1} \right]  \right) \right\}^{3/(2-2l)}
\label{correction}
\end{equation}
and $q=(a/a_*)^2$ with $a_*=\sqrt{\gamma j / 3}l_{\text{Pl}} $. Function~(\ref{correction}) depends on the ambiguity
parameter $l$. As it was shown by Bojowald \cite{Bojowald:2002ny} the value of this parameter is quantised 
according to $l_k=1-(2k)^{-1} \geq 1/2 , \ k \in \mathbb{N} $. For the further investigations we choose the 
representative value $l=3/4$. In the semi-classical region $ a_* \gg a \gg a_i$ 
expression (\ref{correction}) simplify to the form
\begin{equation}
D=D_*a^n 
\end{equation}
where
\begin{equation}
D_*  = \left( \frac{3}{1+l}  \right)^{3/(2-2l)} a_*^{-3(2-l)/(1-l)}  \ \ \text{and} \ \ n = 3(2-l)/(1-l)  \ .
\end{equation}

Now, due to the Hamilton equations we can derive the Friedmann and Raychaudhuri equations for the flat FRW universe filled with a 
homogeneous scalar field
\begin{eqnarray}
H^2 &=& \frac{8\pi G}{3} \left[ \frac{\dot{\phi}^2}{2D} +V(\phi)    \right]  \ ,    \label{Fried1}      \\
\frac{\ddot{a}}{a} &=& -\frac{8\pi G}{3} \left[ \frac{\dot{\phi}^2}{D} \left( 1-\frac{\dot{D}}{4HD}  \right) -V(\phi) \right].   
\label{Raych1}
\end{eqnarray}
The equation of motion for the scalar field with quantum corrections has the form
\begin{equation}
\ddot{\phi}+\left(3H - \frac{\dot{D}}{D} \right)\dot{\phi} + D\frac{dV}{d\phi} =  0.
\label{eom}
\end{equation}
As we mentioned before, for the further investigations we simplify equations (\ref{Fried1}), (\ref{Raych1}) and (\ref{eom})
assuming $V(\phi) = 0$.

The expression for the quantum correction $D$ is complicated and it is impossible to find an analytical 
solution for the equations of motion. In fact we even do not need it for the future investigations.
To calculate the spectrum of gravitons we need to know analytical solutions only 
for the inner and outer states. We choose the $| \text{in} \rangle$ and $| \text{out} \rangle$ states respectively in the 
quantum and classical regimes. The expression for the quantum correction (\ref{correction})
simplifies to the form $D=D_*a^n$ for the $a_i <  a \ll a_*$ and $D=1$ for $a \gg a_* $.
In these limits we can find the analytical solutions for the equations of motion (\ref{Fried1}), (\ref{Raych1}) and (\ref{eom}). 
It is useful to introduce the conformal time $d\tau = dt/a $ to solve equations and for the further investigations.
In the next step we must to fit obtained asymptotic solutions using a global numerical solution.
The solution for the evolution of the scale factor in the quantum limit has the form
\begin{equation}
a=\xi(-\tau+\beta)^p
\label{solution1}
\end{equation}
where $p=2/(4-n)$. The solution in the classical limit we obtain putting simply $l=2$ what gives $D=1$ and $p=1/2$.
The constants of integration $\xi$ and $\beta$ we fix with the use of a numerical solution applying formula
\begin{eqnarray}
\xi =  a|_{\tau} \left[-\frac{a'|_{\tau}}{p \ a|_{\tau} } \right]^p   \ 
 \ \text{and} \ \ \beta  = -\tau-p\frac{a|_{\tau}}{a'|_{\tau}}.
\label{fixing}
\end{eqnarray}
The value of the conformal time $\tau$ must be chosen in a proper way for the given regions. We will discuss
this question in more details later. 

\begin{figure}[ht!]
\centering
\includegraphics[width=7cm,angle=270]{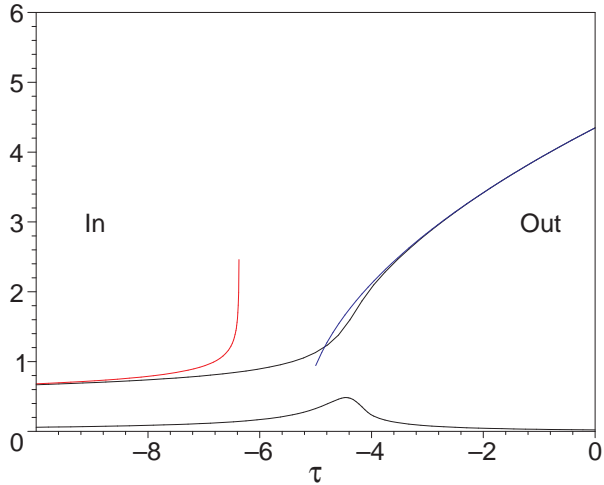}
\caption{Numerically calculated evolution of the scale factor (top black line) and 
        the Hubble parameter (bottom black line) for the model with $j=100$, $l=3/4$ and with 
        initial conditions (\ref{init1}) and (\ref{init2}). 
        The approximated solutions described by (\ref{solution1}) and~(\ref{solution2}).}
\label{fig:solution1}
\end{figure}

Our point of reference is the numerical solution. To make this description complete we must  
choose the proper boundary conditions for the numerical solution. We use here the condition 
for the Hubble radius which must be larger than the limiting value $a_i$ \cite{Lidsey:2004ef},
what gives us 
\begin{equation}
k \simeq |H|a < \frac{a}{a_i} \ \  \Rightarrow \ \  |H|a_i < 1.
\label{condition1}
\end{equation} 
The next condition requires that the scale factor must be greater than 
$a_i$ at the bounce. It is fulfilled taking $a|_{\tau_0} = a_{*}$ for
some value of the conformal time $\tau_{0}$. In fact the conformal time is 
unphysical variable and their value can be chosen arbitrary.
The physical outcomes do not depend on coordinates because the 
theory is invariant under local diffeomorphisms.
So as an example we can choose 
\begin{eqnarray}
a|_{\tau_0=-4} &=&  a_{*}  \label{init1} \\
a'|_{\tau_0=-4} &=&  l_{\text{Pl}}  \label{init2}
\end{eqnarray}
The chosen value of  $a'|_{\tau_0=-4}$ holds the condition (\ref{condition1}). 
Namely for $j=100$ we have $|H_*|a_i = 0.084 < 1$. The numerical solution 
is shown in Fig.~\ref{fig:solution1} as a black line.  
We fix the boundary approximations in $\tau_1=-20$  and $\tau_2=-1$. The the numerical  
solution gives us for these points
\begin{eqnarray}
a|_{\tau_1=-20} &=&  0.536  l_{\text{Pl}} \ , \ a'|_{\tau_1=-20} =  0.007  l_{\text{Pl}} \\
a|_{\tau_2=-1} &=&  3.912  l_{\text{Pl}} \ , \ a'|_{\tau_2=-1} =  0.461  l_{\text{Pl}}.
\end{eqnarray}
Now with the use of expressions (\ref{fixing}) we can fix the approximated solutions.
However when we use formula (\ref{fixing}) directly to calculate parameters in solution (\ref{solution1})
for the outer state we obtain complex $\xi$. It is due to the expression under square $(p=1/2)$ is 
negative. So to put away complex numbers we redefine the exit solution to the form 
\begin{equation}
a=\kappa\sqrt{\tau+\zeta} 
\label{solution2}
\end{equation} 
where 
\begin{eqnarray}
\kappa &=& -i \xi  \\
\zeta &=& -  \beta.
\end{eqnarray}

We show in Fig.~\ref{fig:solution1} how these approximated solutions match with 
solutions obtained numerically. As we can see these approximated solutions well 
describe the evolution in the neighbourhood of $a_*$.

\section{Gravitational waves}

We have already mentioned in section \ref{sec:intro} that gravitational waves
can be abundantly produced during the accelerating phase. In this section
we want to show in details how it works and calculate properties
of produced gravitons. To describe the spectrum of gravitons it 
is common to use the parameter
\begin{equation}
\Omega_{\text{gw}}(\nu) =\frac{\nu}{\rho_c}\frac{d \rho_{\text{gw}}}{d \nu}
\label{omegaGW}
\end{equation}
where $\rho_{\text{gw}} $ is the energy density of gravitational waves 
and $\rho_c$ is present critical energy density. Our goal in this section 
is to calculate the function $\Omega_{\text{gw}}(\nu)$ for the 
gravitons produced during the super-inflationary phase.

The gravitational waves $h_{ij}$ are the perturbations of the background spacetime in the form
\begin{equation}
ds^2=a^2(\tau) \left[ -d\tau^2 +(\delta_{ij}+h_{ij} )dx^i dx^j \right]
\label{metric1}
\end{equation}
where $|h_{ij}|\ll 1$. Using constraints $h^i_i=\nabla_ih^i_j=0 $ we can see 
that tensor $h_{ij}$ have only two independent components $h^1_1=-h^2_2=h_+$ 
and $h^2_1=h^1_2=h_{\times}$. These components correspond to two different 
polarisations of gravitational waves. Inserting the perturbed metric (\ref{metric1}) 
to the Hilbert-Einstein action $S_{\text{H-E}}=(1/16 \pi G) \int d^4x \sqrt{-g}R$ gives the series
 $S_{\text{H-E}}=S^{(0)}+S^{(1)}+S^{(2)}+\dots$, where the second order term have a form 
\begin{equation}
S^{(2)}_t=\frac{1}{64\pi G} \int d^4x a^3 \left[ 
\partial_t h^i_j\partial_t h^j_i-\frac{1}{a^2}\nabla_k h^i_j\nabla_k h^j_i   \right]
 = \frac{1}{32\pi G} \int d^4x a^3 \left[ \dot{h}_{\times}^2+\dot{h}_{+}^2-
\frac{1}{a^2}\left(\vec{\nabla} h_{\times} \right)^2-\frac{1}{a^2}\left(\vec{\nabla} h_{+}\right)^2  \right]
\label{action1}
\end{equation}
and give us the action for the gravitational waves. The two kinds of polarisations are not coupled and can be treated separately.
To normalise the action and simplify the notation it us useful to introduce the variable
\begin{equation}
h=\frac{h_{+}}{\sqrt{16\pi G}}=\frac{h_{\times}}{\sqrt{16\pi G}},
\end{equation}
what leads to the expression for the action in the form
\begin{equation}
S_t = \frac{1}{2} \int d^4x a^3 \left[  \dot{h}^2-\frac{1}{a^2} \left( \vec{\nabla} h \right)^2 \right].
\label{actionh}
\end{equation}
This action is the same like the action for an inhomogeneous scalar field without the potential.
Inverse volume corrections can be therefore introduced in the same way like in the case of the scalar field.
As we mentioned in Introduction there are also holonomy corrections to this action. Here we 
consider however the influence from the better examined inverse volume corrections.    

As it was shown by Mulryne and Nunes \cite{Mulryne:2006cz}, in the context of scalar field perturbations, it is useful
to introduce the variable $u=a D^{-1/2} h $ and rewrite the action~(\ref{actionh}) with quantum corrections to the form
\begin{equation}
S_{\text{t}} = \frac{1}{2}\int d \tau d^3 {\bf x} [ u^{'2}-D \delta^{ij} \partial_i u \partial_j u  -m^2_{\text{eff}}u^2 ]
\end{equation}
where 
\begin{equation}
m^2_{\text{eff}} = - \frac{\sqrt{D}}{a} \left( \frac{a}{\sqrt{D}} \right)^{''} .
\end{equation}
Till now the considerations of the gravitational waves has been purely classical.
The next step is the quantisation of the classical gravitational waves what brings 
us the concept of gravitons. To quantise the field $u$ we need to firstly calculate 
conjugated momenta      
\begin{equation}
\pi(\tau,{\bf x})=\frac{\delta S_{\text{t}}}{\delta u'} = u '. 
\end{equation}
The procedure of quantisation is the simple change of fields $u$ and $\pi$ for the operators just adding hats and
to introduce the relations of commutation. We decompose operators considered for the Fourier modes
\begin{eqnarray}
\hat{u}(\tau,{\bf x} )   &=& \frac{1}{2(2\pi)^{3/2}} \int d^3{\bf k } \left[ \hat{u}_{{\bf k}}(\tau) e^{i{\bf k}\cdot {\bf x}} +  
\hat{u}_{{\bf k}}^{\dagger}(\tau) e^{-i{\bf k}\cdot {\bf x}}  \right],   \label{decomp1}       \\
\hat{\pi}(\tau,{\bf x} ) &=& \frac{1}{2(2\pi)^{3/2}} \int d^3{\bf k }
 \left[ \hat{\pi}_{{\bf k}}(\tau) e^{i{\bf k}\cdot {\bf x}} +  
\hat{\pi}_{{\bf k}}^{\dagger}(\tau) e^{-i{\bf k}\cdot {\bf x}}  \right],  \label{decomp2}        
\end{eqnarray}
where the Fourier components fulfil the relations of commutation 
\begin{eqnarray}
\left[ \hat{u}_{{\bf k}}(\tau) ,\hat{\pi}_{{\bf p}}^{\dagger}(\tau) \right]  &=& i \delta^{(3)}({\bf k} - {\bf p}),\label{com1}\\
\left[ \hat{u}_{{\bf k}}(\tau)^{\dagger} ,\hat{\pi}_{{\bf p}}(\tau) \right]  &=& i \delta^{(3)}({\bf k} - {\bf p}),\label{com2} \\
\left[ \hat{u}_{{\bf k}}(\tau) ,\hat{\pi}_{{\bf p}}(\tau) \right]  &=& i \delta^{(3)}({\bf k} + {\bf p}), \label{com3}  \\
\left[ \hat{u}_{{\bf k}}(\tau)^{\dagger} ,\hat{\pi}_{{\bf p}}^{\dagger}(\tau) \right]  &=& i \delta^{(3)}({\bf k} + {\bf p}).   
\label{com4}
\end{eqnarray}

To express the Fourier modes in terms of the annihilation and creation operators we need to solve 
the quantum Hamilton equations
\begin{eqnarray}
\hat{u}^{'}   &=& i [ \hat{H}_{\text{t}}, \hat{u}   ],  \label{Ham1}  \\
\hat{\pi}^{'} &=& i [ \hat{H}_{\text{t}}, \hat{\pi} ].  \label{Ham2}
\end{eqnarray}

The Hamilton operator have the form  
\begin{eqnarray}
\hat{H}_{\text{t}} &=& \frac{1}{2}\int d^3 {\bf x} [ \hat{\pi}^{2}+D \delta^{ij}  \partial_i \hat{u} \partial_j \hat{u} 
  +m^2_{\text{eff}}\hat{u}^2 ]  \nonumber \\ 
 &=&  \frac{1}{2} \frac{1}{4(2\pi)^{3}}  \int d^3 {\bf x} d^3 {\bf k}  d^3 {\bf q} 
  \left[ \hat{\pi}_{{\bf k}} e^{i{\bf k}\cdot {\bf x}} +  
\hat{\pi}_{{\bf k}}^{\dagger} e^{-i{\bf k}\cdot {\bf x}}  \right] \left[ \hat{\pi}_{{\bf q}} e^{i{\bf q}\cdot {\bf x}} +  
\hat{\pi}_{{\bf q}}^{\dagger} e^{-i{\bf q}\cdot {\bf x}}  \right]   \nonumber \\ 
 &+& D \delta^{ij} i \left[ k_i\hat{u}_{{\bf k}} e^{i{\bf k}\cdot {\bf x}} -  
k_i\hat{u}_{{\bf k}}^{\dagger} e^{-i{\bf k}\cdot {\bf x}}  \right] i\left[q_j \hat{u}_{{\bf q}} e^{i{\bf q}\cdot {\bf x}} -  
q_j\hat{u}_{{\bf q}}^{\dagger} e^{-i{\bf q}\cdot {\bf x}}  \right] \nonumber \\ 
 &+& m^2_{\text{eff}}  \left[ \hat{u}_{{\bf k}} e^{i{\bf k}\cdot {\bf x}} +  
\hat{u}_{{\bf k}}^{\dagger} e^{-i{\bf k}\cdot {\bf x}}  \right]  \left[ \hat{u}_{{\bf q}} e^{i{\bf q}\cdot {\bf x}} +  
\hat{u}_{{\bf q}}^{\dagger} e^{-i{\bf q}\cdot {\bf x}}  \right]    \nonumber \\ 
          &=& \frac{1}{4} \int d^3 {\bf k}\left[  \hat{\pi}_{{\bf k}} \hat{\pi}_{{\bf k}}^{\dagger}+
 \hat{\pi}_{{\bf k}}^{\dagger} \hat{\pi}_{{\bf k}} +\left(Dk^2 + m^2_{\text{eff} }\right)
\left( \hat{u}_{{\bf k}} \hat{u}_{{\bf k}}^{\dagger}+ \hat{u}_{{\bf k}}^{\dagger} \hat{u}_{{\bf k}}  \right)      \right] 
\label{Ham}
\end{eqnarray}
where we inserted decompositions (\ref{decomp1}) and (\ref{decomp2}). When we apply the Hamiltonian~(\ref{Ham}) and the 
decompositions~(\ref{decomp1}) and~(\ref{decomp2}), the Hamilton equations~(\ref{Ham1}) and~(\ref{Ham2}) take the forms 
\begin{eqnarray}
\hat{u}^{'}_{{\bf k}}  &=& \hat{\pi} _{{\bf k}},  \label{Ham11}   \\
\hat{\pi}^{'}_{{\bf k}} &=& -\left(Dk^2 + m^2_{\text{eff} }\right) \hat{u}_{{\bf k}}.  \label{Ham22} 
\end{eqnarray}
The general solution of these equations has the form
\begin{eqnarray}
\hat{u}_{ {\bf k}  }(\tau)  &=&  \hat{a}_{{\bf k} } f(k,\tau)+ 
\hat{a}_{-{\bf k} }^{\dagger} f^{*}(k,\tau), \label{sol11}   \\
\hat{\pi}_{{\bf k} }(\tau)  &=&  \hat{a}_{{\bf k} } g(k,\tau)+ 
\hat{a}_{-{\bf k} }^{\dagger} g^{*}(k,\tau). \label{sol22}
\end{eqnarray}
where $f(k,\tau)'=g(k,\tau)$.
When we insert these solutions to the Fourier decompositions~(\ref{decomp1}) and~(\ref{decomp2}) we simply obtain
\begin{eqnarray}
\hat{u}(\tau,{\bf x} )   &=& \frac{1}{(2\pi)^{3/2}} \int d^3{\bf k } \left[ f(k,\tau)  \hat{a}_{{\bf k}}
e^{i{\bf k}\cdot {\bf x}} +  
f^*(k,\tau) \hat{a}_{{\bf k}}^{\dagger} e^{-i{\bf k}\cdot {\bf x}}  \right]   \label{decomp11} \ ,      \\
\hat{\pi}(\tau,{\bf x} ) &=& \frac{1}{(2\pi)^{3/2}} \int d^3{\bf k } \left[ g(k,\tau)
\hat{a}_{{\bf k}} e^{i{\bf k}\cdot {\bf x}} +  g^*(k,\tau) \hat{a}_{{\bf k}}^{\dagger} e^{-i{\bf k}\cdot {\bf x}}  \right]. 
\label{decomp22}        
\end{eqnarray}
The mode functions fulfils the so called Wronskian condition
\begin{equation}
f^*(k,\tau) g(k,\tau) - f(k,\tau) g^*(k,\tau)=-i \label{Wronskian}
\end{equation}
as a result of relations of commutation (\ref{com1}-\ref{com4}). These relations is important to 
normalise properly the mode functions.
 
The Hamilton equations~(\ref{Ham11}) and~(\ref{Ham22}) together with (\ref{sol11}) give us the equation for the mode function
\begin{equation}
\frac{d^2}{d\tau^2}f(k,\tau) + \left[ D k^2 +m^2_{\text{eff}} \right] f(k,\tau) = 0. \label{modeeq}
\end{equation}
This equations has two regimes. The first one called adiabatic corresponds to the situation when
$D k^2 +m^2_{\text{eff}} \equiv \Gamma  \gg 0 $. The second one leads to the 
super-adiabatic amplification and corresponds to the situation when $\Gamma  \ll 0$.
The creation of the gravitational waves corresponds to the case of the 
super-adiabatic amplification. 

\begin{figure}[ht!]
\centering
$\begin{array}{cc}   
\includegraphics[width=6cm,angle=270]{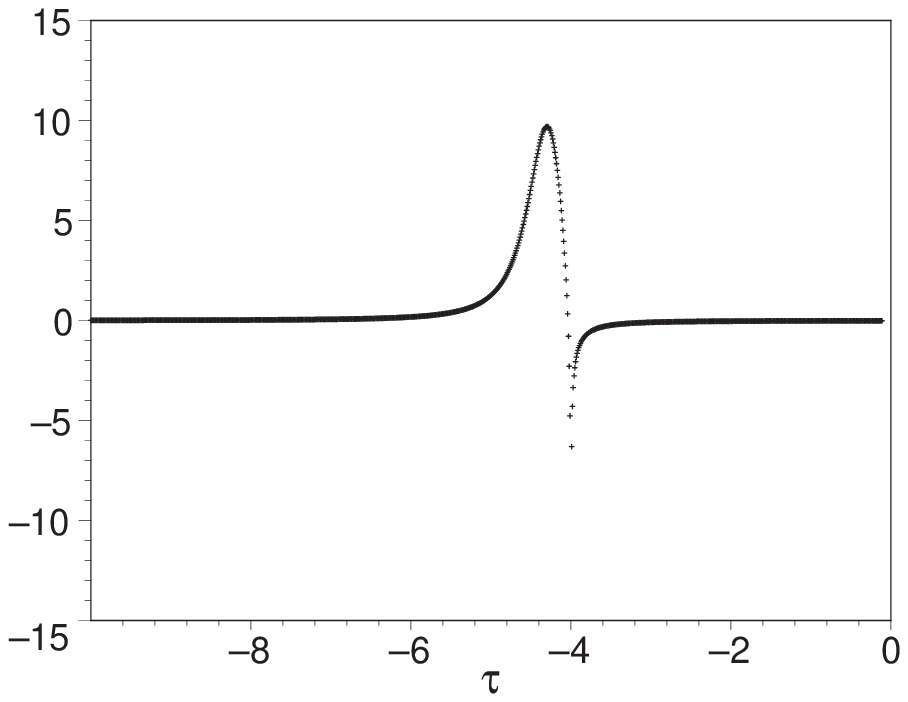}  &  \includegraphics[width=6cm,angle=270]{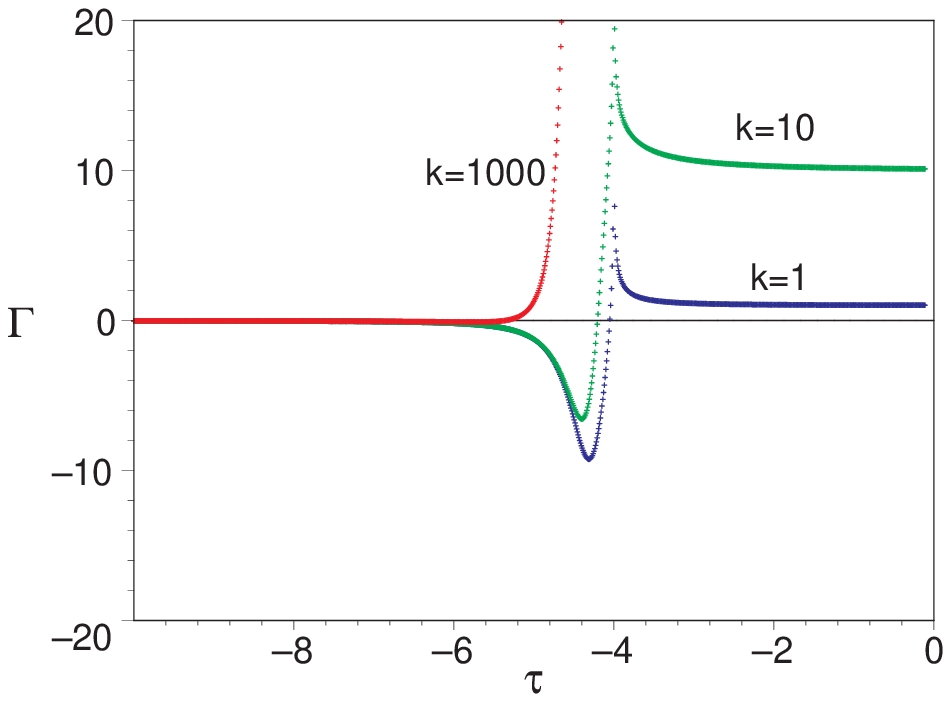}   
\end{array}
$\caption{ 
 {\bf Left }: Evolution of the parameter $-m^{2}_{\text{eff}}$ for the model considered. 
 { \bf Right }: Evolution of the function $\Gamma=D k^2 +m^2_{\text{eff}}$ for different 
values of $k$. }
\label{pumpfield}
\end{figure}

We can now investigate which modes are amplified. The condition $\Gamma  \ll 0$ corresponds to $D k^2 \ll -m^2_{\text{eff}}$.
In Fig.~\ref{pumpfield} we see the evolution of $-m^2_{\text{eff}}$. As we see the condition 
for the creation of gravitational waves is fulfilled in the region of the super-inflation $( \tau \in [ \sim -6 ,  \sim -4 ] )$ . 
In the right panel we can see the evolution of $\Gamma$ for three different values of $k$.

The value of $k$ is dimensionless in the undertaken scheme. To obtain a dimensional value we must 
multiply it simply by the corresponding scale factor which has a dimension of length. 
The wave number $k$, for example for the final state $a_{\text{f}}$, corresponds to the frequency
\begin{equation}
\nu = \frac{k}{2\pi a_{\text{f}}} \left( \frac{a_{\text{f}}}{a_{\text{today}}} \right) \label{freq}
\end{equation} 
measured today. The important task is to calculate the factor $ a_{\text{today}}/a_{\text{f}} $. We can make 
the decomposition 
\begin{equation}
\frac{a_{\text{today}}}{a_{\text{f}}} = 
\frac{a_{\text{today}}}{a_{\text{dec}}} \frac{a_{\text{dec}}}{a_{\text{rh}}} \frac{a_{\text{rh}}}{a_{\text{i-end}}} 
\frac{a_{\text{i-end}}}{a_{\text{i-star}}} \frac{a_{\text{i-start}}}{a_{\text{f}}}
\end{equation}
where 
\begin{eqnarray}
a_{\text{dec}} &\rightarrow& \text{ photons decoupling} , z _{\text{dec}}\simeq 1070 ,
 T_{\text{dec}} \simeq 3\cdot 10^3 \ \text{K} 
=   0.2 \ \text{eV}  \nonumber \\
a_{\text{rh}}  &\rightarrow&   \text{ reheating phase  }  , T_{\text{rh}} = T_{\text{GUT}} \simeq 10^{14} \ \text{GeV}\nonumber \\
a_{\text{i-end}}  &\rightarrow& \text{end of inflation  }       \nonumber  \\
a_{\text{i-start}}  &\rightarrow& \text{beginning of inflation  }    \nonumber
\end{eqnarray}
We can assume the sudden reheating approximation $(a_{\text{rh}} \simeq a_{\text{i-end}} )$ and the standard value of e-folding 
number for inflation $N \equiv \ln ( {a_{\text{i-end}}}/{a_{\text{i-star}}} ) = 63 $. Then increase of the scale factor forms
the final state till present value assumes 
\begin{equation}
\frac{a_{\text{today}}}{a_{\text{f}}} \simeq (1+z _{\text{dec}}) \cdot \frac{ T_{\text{GUT}} }{ T_{\text{dec}}  } \cdot 
  e^{N} \cdot   \frac{a_{\text{i-start}}}{a_{\text{f}}} \simeq 10^{56}. 
\end{equation}
We had assumed here that ${a_{\text{i-start}}}/{a_{\text{f}}}\simeq 10^2 $, this value can be obtained from numerical 
simulations like these in Ref.~\cite{Tsujikawa:2003vr}. Now we can return to equation (\ref{freq}) and then calculate 
the maximal frequency of produced gravitons. From the right panel in Fig.~\ref{pumpfield} we can see that 
in the vicinity of $\tau \sim -6$ we have the transition from positive to negative values of $\Gamma$. So as we can see for some 
$10<k_{\text{k}}<1000$ we have the transition from the adiabatic to the super-adiabatic regime. From the numerical
investigation we obtain $k_{\text{tr}} \simeq 500$. In fact the function $\Gamma$, for small values of $\tau$, is  always 
slightly below zero even for greater values of $k$ than $k_{\text{tr}}$. But we assume that this effect is negligible.
In fact higher values of $k$ easily reach the transplanckian scales and it is not clear that we should trust the 
standard physics in this regime. The wave number $k_{\text{tr}}$ corresponds roughly to the scales $a_*$. So the value of
$k_{\text{tr}}$ corresponds to $ k_{\text{f}} = (a_*/a_{\text{f}})   k_{\text{tr}} \simeq 250 $ in the final state 
$a_{\text{f}}(\tau=-1)=3.912  l_{\text{Pl}}$.  Applying equation~(\ref{freq}) we obtain the maximal frequency for the present epoch
$\nu_{\text{max}} \simeq 2 \cdot 10^{-12} \ \text{Hz}$.
This value corresponds to the scales $\lambda_{\text{min}} \simeq 5 \ \text{kpc}$. It is instructive to consider the model
without inflation. In this situation the maximal frequency of the relic gravitons would be
$\nu_{\text{max}} \simeq 2 \cdot 10^{15} \ \text{Hz}$ what is extremely huge number. In fact it is possible 
that GUT energy scale and inflation cover each other in some place. It was somehow one of the motivation
to introduce inflation to solve the problem of topological defects. In the case when the GUT scale occurs after
thee reheating the problem of topological defects must be solved in a different way. We mention this problem
to show that it is possible that the value ${a_{\text{today}}}/{a_{\text{f}}}$ can be lower than calculated before.

After this analysis we can return now to equation~(\ref{modeeq}). 
We need to use the definition of the quantum correction $D$ and effective mass $m^2_{\text{eff}}$ in the quantum regime.
It is also useful to rescale the conformal time and introduce $-\eta=-\tau+\beta$.
We will back later to the previous definition because an additional degree of 
freedom $\beta$ is necessary to fit properly the boundary solutions 
with numerical one. Equation~(\ref{modeeq}) takes the form 
\begin{equation}
\frac{d^2}{d\eta^2}f(k,\eta)+\left[ D_*\xi^n(-\eta)^{np}k^2-p(p-1) \right]f(k,\eta) = 0
\end{equation}
and the general solution in terms of Bessel functions have the form
\begin{equation}
f (k,\eta)= C_1\sqrt{-\eta}J_{|\nu|}(x)+ C_2\sqrt{-\eta}Y_{|\nu|}(x) \label{solmodes}
\end{equation}
with
\begin{eqnarray}
 x   &=&  k \frac{2\sqrt{D_*\xi^n}}{|2+np|} (-\eta)^{(2+np)/2}, \\
 \nu &=& -\frac{\sqrt{1+4p(p-1)}}{2+np}.
\end{eqnarray}
With the use of the Wronskian condition  (\ref{Wronskian}) we can rewrite the solution (\ref{solmodes}) to the form
\begin{equation}
f (k,\eta)  =  \sqrt{\frac{\pi}{2|2+np|}}\sqrt{-\eta}\left[ D_1 H^{(1)}_{|\nu|}(x) + D_2 H^{(2)}_{|\nu|}(x)  \right] 
\label{solmodes2}
\end{equation}
where we introduced Hankel functions defined as 
\begin{eqnarray}
H^{(1)}_{|\nu|}(x) &=& J_{|\nu|}(x) +i Y_{|\nu|}(x) \\
H^{(2)}_{|\nu|}(x) &=& J_{|\nu|}(x) -i Y_{|\nu|}(x).
\end{eqnarray}
and the constants $D_1$ and $D_2$ enjoy the relation $|D_1|^2-|D_2|^2=1$. To fix values of the constants $D_1$ and $D_2$
we must consider the high energy limit, namely $x \gg 1 $. In this limit the Bessel functions behave as follow 
\begin{eqnarray}
J_{|\nu|}(x) \rightarrow  \sqrt{\frac{2}{\pi x}} \sin \left(x- \frac{|\nu|\pi}{2} -\frac{\pi}{4} \right), \\
Y_{|\nu|}(x) \rightarrow  \sqrt{\frac{2}{\pi x}} \cos \left(x- \frac{|\nu|\pi}{2} -\frac{\pi}{4} \right),
\end{eqnarray}
what give us  $H^{(1)}_{|\nu|}(x) \rightarrow \sqrt{{2}/{(\pi x)}} \exp{\left[ix - i{|\nu|\pi}/{2} -i{\pi}/{4} \right]}$ and
$H^{(2)}_{|\nu|}(x) \rightarrow \sqrt{{2}/{(\pi x)}} \exp{\left[ -ix +i{|\nu|\pi}/{2} +i{\pi}/{4} \right]}$. Classically 
the limit  $x \gg 1 $ corresponds to advanced solution called the Bunch-Davies vacuum  $e^{-ik\tau}/\sqrt{2k}$. Generally 
the limit obtained here differs from the classical one but can be restored taking $l=2$. Then to obtain the proper high energy limit 
we must choose $D_2=0$ and $D_1=\exp{\left[ i{|\nu|\pi}/{2} +i{\pi}/{4} \right]}$. Applying evaluated values of $D_1$ and $D_2$
to the solution~(\ref{solmodes2}) we finally obtain mode functions for the initial state
\begin{eqnarray}
f_i(k,\tau)  &=& \mathcal{N}  \frac{1}{\sqrt{k}}\sqrt{-k\tau+k\beta} H^{(1)}_{|\nu|}(x)  \\
g_i(k,\tau) &=& \mathcal{N}  \sqrt{k}\sqrt{-k\tau+k\beta}
\left[ -\frac{1}{2} \frac{ H^{(1)}_{|\nu|}(x) }{-k\tau+k\beta } -\frac{2+np}{|2+np|}\sqrt{D_*\xi^n}(-\tau+\beta)^{\frac{np}{2}}
 \left( \frac{|\nu|}{x}H^{(1)}_{|\nu|}(x)-H^{(1)}_{|\nu|+1}(x)     \right)  \right]
\end{eqnarray}
where 
\begin{equation}
\mathcal{N}=e^{i\left( \frac{|\nu|\pi}{2} +\frac{\pi}{4}  \right)}  \sqrt{\frac{\pi}{2|2+np|}}.
\end{equation}
The functions $g_i(k,\tau)$ were calculated from relation $f(k,\tau)'=g(k,\tau)$ showed earlier. We had used also
the expression for derivative of the Hankel functions in the form
\begin{equation}
\frac{dH^{(1)}_{|\nu|}(x)}{dx} =\frac{|\nu|}{x}H^{(1)}_{|\nu|}(x)-H^{(1)}_{|\nu|+1}(x).    
\end{equation}
Similar investigations lead to the expression for the mode functions for the final state.  
We must remember that it is however not only a simple set $l=2$ but also the change of the solution
for the scale factor to this expressed by~(\ref{solution2}). In this case we obtain
\begin{eqnarray}
f_f(k,\tau) &=& e^{i\frac{\pi}{4}}  \sqrt{\frac{\pi}{4k}} \ \sqrt{k\tau+k\zeta}H^{(1)}_0(k\tau+k\zeta),   \\
g_f(k,\tau) &=& e^{i\frac{\pi}{4}} \sqrt{\frac{\pi k}{4}}\sqrt{k\tau+k\zeta}
\left[ \frac{1}{2} \frac{H^{(1)}_0(k\tau+k\zeta)}{k\tau+k\zeta} -  H^{(1)}_1(k\tau+k\zeta)  \right].
\end{eqnarray}

\begin{figure}[ht!]
\centering
\includegraphics[width=7cm,angle=270]{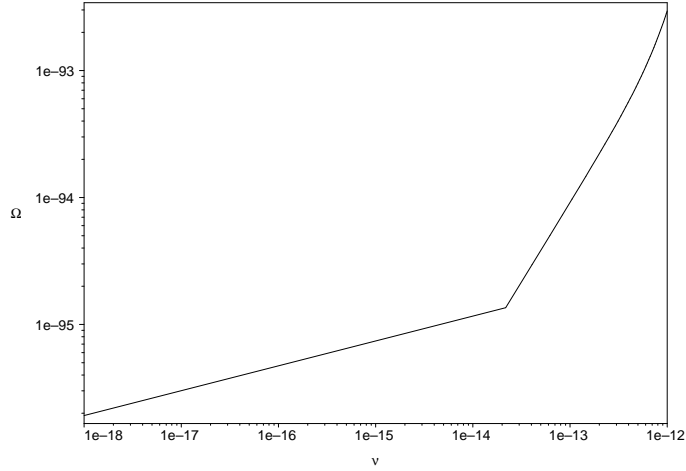}
\caption{Spectrum of relic gravitons for the model with $j=100$ and $l=3/4$. Frequency scale in Hertz.}
\label{spect1}
\end{figure}

Now we are ready to consider the creation of the gravitons during the transition from some initial to final states. 
The initial vacuum state $| 0_{\text{in}}\rangle$  is determined by 
$\hat{a}_{\text{k}}| 0_{\text{in}}\rangle = 0$, where  $\hat{a}_{\text{k}}$ is the initial annihilation operator for $\tau_i$. 
The relation between annihilation and creation operators for the initial and final states is given 
by the Bogoliubov transformation      
\begin{eqnarray}
\hat{b}_{{\bf k}} &=& B_{+}(k) \hat{a}_{{\bf k}} + B_{-}(k)^{*}  \hat{a}_{-{\bf k}}^{\dagger} \ , \label{Bog1}  \\
\hat{b}_{{\bf k}}^{\dagger} &=& B_{+}(k)^{*}\hat{a}_{{\bf k}}^{\dagger} + B_{-}(k)    \hat{a}_{-{\bf k}} \label{Bog2} 
\end{eqnarray}
where $|B_{+}|^2-|B_{-}|^2=1$. Because we are working in the Heisenberg description the vacuum state does not 
change during the evolution. It results that
$\hat{b}_{{\bf k}}| 0_{\text{in}}\rangle=B_{-}(k)^{*}  \hat{a}_{-{\bf k}}^{\dagger}| 0_{\text{in}}\rangle $  is 
differ from zero when $B_{-}(k)^{*}$ is a nonzero function. This means that in the final state  
graviton field considered is no more in the vacuum state without particles. The number of produced particles in the 
final state is given by  
\begin{equation}
\bar{n}_{{\bf k}} = \frac{1}{2} \langle 0_{\text{in}} |\left[ \hat{b}_{{\bf k}}^{\dagger}\hat{b}_{{\bf k}}+
 \hat{b}_{-{\bf k}}^{\dagger}\hat{b}_{-{\bf k}} \right]| 0_{\text{in}} \rangle =|B_{-}(k)|^2. \label{particles}
\end{equation}
Using relations~(\ref{sol11})and (\ref{sol22}) and the Bogoliubov transformation (\ref{Bog1}) and (\ref{Bog2}) we obtain
\begin{eqnarray}
B_{-}(k)&=&
\frac{f_i(k,\tau_i)g_f(k,\tau_f) -g_i(k,\tau_i) f_f(k,\tau_f)}{f_f^*(k,\tau_f)g_f(k,\tau_f)-g_f^*(k,\tau_f)f_f(k,\tau_f)} 
 = i\left[  f_i(k,\tau_i)g_f(k,\tau_f) -g_i(k,\tau_i) f_f(k,\tau_f)    \right]  \ ,  \\
B_{+}(k)&=& 
\frac{f_i(k,\tau_i)g_f^*(k,\tau_f) -g_i(k,\tau_i) f_f^*(k,\tau_f)}{f_f(k,\tau_f)g_f^*(k,\tau_f)-g_f(k,\tau_f)f_f^*(k,\tau_f)}
 =-i \left[ f_i(k,\tau_i)g_f^*(k,\tau_f) -g_i(k,\tau_i) f_f^*(k,\tau_f)  \right] 
\end{eqnarray}
where simplifications come from the Wronskian condition (\ref{Wronskian}). In the calculations we set 
$\tau_{{i}}=\tau_1=-20$ and $\tau_{{f}}=\tau_2=-1$. These boundaries fully cover the region of gravitational waves creation. 
The energy density of gravitons is given by   
\begin{equation}
d\rho_{\text{gw}} = 2 \cdot \hslash \omega \cdot  \frac{4 \pi \omega^2   d\omega}{(2\pi c)^3} \cdot |B_{-}(k)|^2.
\end{equation}
where we used definition (\ref{particles}). The expression for the parameter $\Omega_{\text{gw}}$ defined by (\ref{omegaGW})
takes now the form
\begin{equation}
\Omega_{\text{gw}}(\nu) =\Omega_0  \cdot \nu^4 \cdot \bar{n}\left[ k = \nu \cdot  2\pi a_{\text{f}} \cdot  
 \left( \frac{a_\text{today}}{a_{\text{f}}} \right)  \right]
\label{omegaGW1}
\end{equation}
where 
\begin{equation}
\Omega_0 = \frac{\hslash c}{c^4}\frac{16\pi^2}{\rho_c} = 
\frac{ 16 \pi^2  \cdot 197.3 \cdot 10^{-15}[\text{MeV} \cdot
 \text{m}]}{3^4 \cdot 10^{32}[\text{m}^4/\text{s}^4] 1.05 \cdot 10^{-5} \cdot h^2_0 
 [\text{GeV}/\text{cm}^3]   }  = 3.66 \cdot h^{-2}_0 \cdot 10^{-49} \ [\text{Hz}^{-4}].
\end{equation}
In the calculations we set present value of the Hubble factor for $h_{0}=0.7$.
In Fig.~\ref{spect1} we show spectrum calculated with formula (\ref{omegaGW1}). The obtained spectrum is extremely weak
in the present epoch. The reason of this tiny amount of the background gravitons is the presence of the standard inflationary phase. 
The super-inflationary phase is placed before the inflation so the energy of gravitons decreases about $10^{27}$ times
during this further phase. 
To see better how presence on the inflation affect this spectrum we show
in Fig.~\ref{spect1} the spectrum of relic gravitons in the model without the inflationary phase.

\begin{figure}[ht!]
\centering
\includegraphics[width=7cm,angle=270]{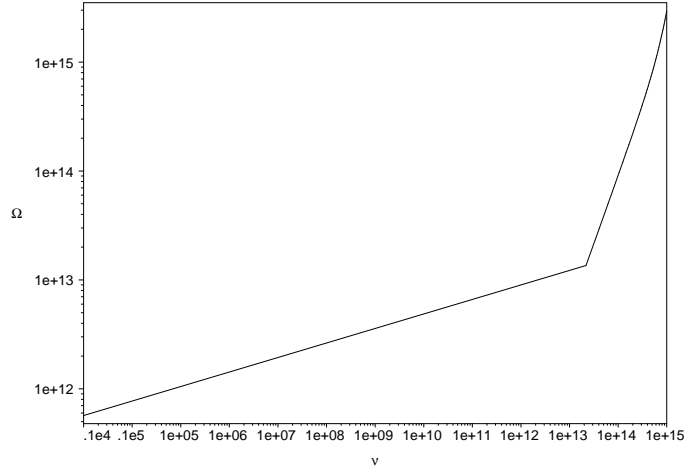}
\caption{Spectrum of relic gravitons for the model without inflation. Frequency scale in Hertz.}
\label{spect2}
\end{figure}

It is clear that in such a model amount of relic gravitons would be extremely large.
As we mentioned before it is possible that GUT energy scales cover partially with the inflation.
In this situation the present amount of the relic gravitons would be higher than this 
in Fig.~\ref{spect1} and reaches to higher energies. For the present state we do not know the 
duration of the inflationary phase exactly. To estimate this value it is necessary to measure the spectrum
of both scalar and tensor parts of primordial fluctuations produced during the standard inflation. 
For the present day we know only the contribution from the scalar part. The further 
generation of CMB telescopes in needful to improve the knowledge the properties of inflation. From our 
calculation we can see however that the inflationary phase is necessary. Without the 
inflation after the super-inflation the present amount of gravitons 
would be easily in reach of present observational skills. From this 
point of view we have found the next motivation to support the 
inflationary model.

\section{Summary}

In summary, we have calculated the spectrum of gravitons produced during the super-inflationary phase
induced by Loop Quantum Gravity effects. In the calculations we considered inverse-volume 
corrections to the dynamics and to the equation for the tensor modes. We have solved analytically 
equation for the tensor modes in the quantum and classical regimes. Both solutions 
we had matched by numerical solution for background dynamics. We have obtained 
spectra of relic gravitons for the models considered. In the first model we assumed
the presence of the inflationary phase after the super-inflation. In that case we obtained
presently a negligible amount of relic gravitons. However in the second model 
without inflation the present amount of graviton background would be 
unnaturally high. This results state that the period of inflation after the super-inflation
is necessary to avoid the problem of relic gravitons. Nowadays the main motivation 
to introduce the inflationary phase comes from ability to creation of fluctuations.
Our investigations based on loop quantum cosmology support the inflationary model.   
   
Results obtained in this work differ from our previous investigations
\cite{Mielczarek:2007zy}. The difference comes mainly from
different assumptions about the evolution of the Universe after the super-inflation.
The previous results correspond to the situation when inflation
and GUT energy scales cover each other rather than to the case of very short 
inflation. In this paper we have considered both models agreeing with 
the present paradigm and its modification.

\begin{acknowledgments}
This work was supported in part by the Marie Curie Actions Transfer of
Knowledge project COCOS (contract MTKD-CT-2004-517186).
\end{acknowledgments}

\end{document}